Max Schuchard* and Nicholas Hopper*

# E-Embargoes: Discouraging the Deployment of Traffic Manipulating Boxes With Economic Incentives

## Abstract

An increasing number of systems have been proposed or deployed to the transit core of the Internet with the goal of observing and manipulating traffic in flight, systems we term *Traffic Manipulating Boxes*. Examples of these include: decoy routing systems, surveillance infrastructure like the NSA's alleged QUANTUM project, and traffic shaping middleboxes. In this work, we examine a new approach that a *routing capable adversary* might take to resisting these systems: the use of economic pressure to incentivize ISPs to remove them. Rather than directly attacking the availability of these systems, our attack inflicts economic losses, in the form of reduced transit revenue, on ISPs that deploy them, while at the same time incentivizing ISPs that do not.

We alter and expand upon previous *routing around decoys* attack of Schuchard *et al.*, by adjusting the priority given to avoiding TMBs. This reduces or eliminates the key costs faced by routing capable adversary while maintaining the effectiveness of the attack. Additionally, we show that since the flow of traffic on the Internet is directly related to the flow of cash between ISPs, a routing capable adversary is actually a powerful *economic adversary*. Our findings show that by preferentially using routes which are free of TMBs, some routing capable adversaries can inflict in excess of a billion dollars in annual revenue losses.

## 1 Introduction

An increasing number of systems that manipulate higher layers of network functionality from positions in the middle of the network, located within transit providers, have been proposed, designed, and deployed. We term these devices *Traffic Manipulating Boxes*, or TMBs. While some TMBs exist to provide simple improvements to the transit functionality of the network, others such as Decoy Routing [22, 26, 39], traffic shaping [3, 40], and active surveillance boxes such as the NSA's alleged QUANTUM-INSERT project [8, 34, 38] seek to use privileged position in the Internet to observe and manipulate traffic in a manner not intended by either its sender or the autonomous system (AS) that hosts its sender.

The question becomes, what can an AS, or even a whole nation, that does not want its traffic to be manipulated by specific TMBs, entities we call *resistors*, do to fight back? As shown by Schuchard *et al.* [35], large ASes and nation-states can use their capability to select which paths they utilize, and which they advertise to others, to launch a straightforward attack on the traffic coverage of TMBs by simply routing their traffic around deployers – ASes that have elected to deploy those boxes – and refusing to send traffic to any destination reachable only through deployers. This attack, termed the RAD (Routing Around Decoys) attack, defeats TMBs since they can neither observe nor manipulate traffic that does not cross them. Recently Houmansadr et al. [23] pointed out that ASes launching a RAD attack incur added costs, possibly making the RAD attack prohibitively expensive.

In this work, we consider how economic forces can work *for* the resistor, rather than against it. Instead of attempting to directly attack the availability of TMBs, our resistors inflict economic damage on deployers, while at the same time incentivizing ASes which elect to not become deployers. Assuming transit ASes are rational economic entities, these incentives will cause deployers to remove TMBs, accomplishing the resistor's goal of minimizing exposure to TMBs. This alternative approach to attacking availability has the added benefit of not requiring the resistors to disconnect themselves from portions of the Internet. In order for this strategy to be viable we must accomplish two distinct goals.

First, we must adjust how the RAD attack functions in an effort to remove the costs to the resistor. The original RAD attack had no concern for the resistor's costs; namely, added infrastructure costs from turning customer ASes into transit ASes, using paths learned from providers over those from customers, and increased path lengths. We present three alterations to the RAD attack which adjust how the attack is executed to successively eliminate each of these costs. This re-

***Corresponding Author: Max Schuchard:** University of Minnesota, E-mail: schuch@cs.umn.edu

***Corresponding Author: Nicholas Hopper:** University of Minnesota, E-mail: hoppernj@umn.edu

duction in resistor costs is not achieved for free, as each of these adjustments reduces the ability of the resistor to find paths free of TMBs. We evaluate these altered RAD attacks, along with the original RAD attack of Schuchard *et al.*, launched from four different resistors, against several deployments, including the deployment proposed to defeat the RAD attack. We find that our lower cost RAD attacks are still able to find paths free of TMBs, and therefore able to migrate a non-trivial amount of traffic away from deployer ASes.

Secondly, we need to maximize the economic costs the resistor can inflict *on the deployers*. Since TMBs require a privileged position on the Internet, deployers need to be large transit ISPs. These ISPs generate revenue by carrying network traffic, the exact thing a routing capable adversary can manipulate. We propose a new form of RAD attack with the goal of coercing deployers to remove TMBs by inflicting economic loss. This attack differs from the original RAD attack, in that it does not *directly* reduce the exposure to the TMBs themselves. This style of attack has a desirable property: the resistor does not have to cut itself off from any destinations on the Internet, it can instead tolerate TMBs functioning while economic pressure builds. In order to evaluate this, we develop a model of Internet scale traffic flow, and build an approximation between our simulated traffic and real world ISP revenue. Our results reveal that routing capable adversaries are extremely powerful economic entities, the largest of which are capable of inflicting a billion dollars in lost annual revenue simply by making different routing decisions.

**Contributions.** We make three primary contributions to the literature on Decoy Routing and Network Interference:

– We demonstrate new, reduced cost "Routing Around Deployers (RAD)" attacks. Compared to the attacks in [35], our attacks do not suffer the costs demonstrated by Houmansadr *et al.* [23], while retaining the ability to find paths free of TMBs.
– We introduce a new adversary model for resisting the deployment of TMBs, the *economic adversary*. We show that the economic adversary can impose significant losses in traffic revenue and leverage *opportunity costs* to discourage targeted deployments of TMBs.
– Within the context of the economic routing adversary, we introduce a new RAD attack, the "reverse poisoning attack," which significantly increases the losses suffered by deployers of TMBs in a targeted deployment by diverting in-bound traffic to the resistor (and its customer ASes) away from deployers.

The rest of this paper is laid out as follows. Section 2 will provide relevant background on TMBs, BGP, the RAD attack, and its costs. Section 3 will present details of our simulation methodology, example resistors, and TMB deployment models. Section 4 will examine how the original RAD attack can be altered to take resistor costs into account. Lastly, in Section 5 we will develop a new type of RAD attack where we harness the economic power of routing capable adversaries to incentivize deployers to remove TMBs from their network.

# 2 Background

## 2.1 Traffic Manipulating Boxes

Traffic manipulating boxes, or **TMBs**, are middle boxes deployed to Autonomous Systems (ASes) that make up the transit core of the Internet. These devices utilize privileges in the transit core to observe or manipulate traffic *not* originated by the owner of the middle box. As covered in the introduction, there are several examples, both proposed and deployed, of TMBs, including decoy routing [26], surveillance devices which actively manipulate traffic in flight [8, 34, 38], and traffic shaping boxes [3, 40].

We refer to the organization that actually wants the TMBs placed on the Internet as the **originator**. The originator places TMBs inside the infrastructure of willing or compelled ISPs, which we refer to as **deployers**. It is of course possible for the originator and deployer to be the same organization; however, this is often not the case. For example, in the case of decoy routing, an outside organization concerned with providing censorship circumvention would be the originator, not the ISPs hosting the Decoy Routers.

Originators have several costs, for example the price of hardware, which they have to pay on a per deployer basis. This means that they have to attempt to minimize the number of deployers while still maximizing the amount of traffic that the TMBs can manipulate. In order to do this, the originator needs to select deployers that will see not only traffic addressed to themselves, but other ASes as well, hence the requirement that TMBs be placed in the transit core of the Internet. Additionally, originators will also have to defray economic losses that deployers suffer as a result of installing TMBs. These side payments would resemble those allegedly provided by the NSA to RSA Labs for the deployment of specific technologies [28]. These side payments will be covered in detail in Section 5.

## 2.2 Routing Capable Adversaries

Some entities – in our context, ASes or coalitions of ASes under the coordination of a nation-state – might not want their traffic exposed to TMBs; we term such entities **resistors**. Resistors can achieve their goal by making sure their traffic does not traverse deployer ASes. Schuchard *et al.* [35] consider one such entity, a censoring nation-state which wished to avoid giving its citizens access to a decoy routing system. They introduced the concept of a *routing capable adversary*, an adversary who could make inter-AS routing decisions based on the presence or absence of TMBs. For ease of terminology, we call any path that crosses one or more TMBs a **tainted path**. Conversely, any path which crosses no TMBs is called a **clean path**.

The routing of traffic between different ASes is handled by the Border Gateway Protocol [31], referred to for the rest of the paper simply as BGP. BGP is a path vector routing algorithm with policies. Essentially, BGP works like a standard path vector routing algorithm; routers receive paths from their neighbors to a given destination, the "best" path is selected, and it is then forwarded out to other neighbors. Where BGP differs from standard path vector routing is in what constitutes a best path and which neighbors a router will forward that path to. In the case of BGP, these decisions are impacted by the business relationship between the AS and its neighbors.

In the attack of Schuchard *et al.*, which we refer to as a **RAD attack** for the rest of the paper, the resistor takes advantage of the fact that most ASes have multiple paths to a given destination. The RAD attack forces ASes to alter standard BGP behavior, at times in contradiction to business policies, in an effort to always select routes free of TMBs. In this specific example, the censoring nation could defeat the decoy routing system, since users could not find a way to have their traffic observed by the decoy routers. The RAD attack does require the resistor to know which ASes are deployers. This is achievable if the resistor can reliably "flag" traffic as having been manipulated by TMBs, allowing a path intersection attack to reliably locate the TMBs; see the original work [35] for details. Section 6 contains a further discussion of locating TMBs.

The ability to avoid TMBs comes at a cost for a routing capable adversary. First, if it can not find a clean path to a destination, the routing capable adversary must cut off connectivity with that destination. Second, the attack, as described originally, in some cases causes increased financial costs for ASes, would require fringe ASes to become transit ASes, increases the load placed on certain links in the topology, and can cause a general degradation of path quality. Many of these added costs placed on the resistor were investigated by Houmansadr et al. [23]. Both the loss of connectivity and these additional costs are dependent on where in the Internet infrastructure TMBs are deployed.

# 3 Methodology, Resistors, & Deployment Models

**Routing Model.** To evaluate both the routing capabilities of new resistor strategies and the economic impacts of the resistor actions, we constructed a BGP simulator [1] similar to those used in prior work [23, 35]. Our simulator is essentially a collection of software routers running BGP. For scalability reasons, and for consistency with prior work, we ignore the internal topology of each AS and treat them as a single BGP speaker. This is acceptable since our strategies are agnostic to intra-AS concerns (e.g. multi-exit discriminators and path origin). Each of these routers faithfully executes the full BGP decision making process: ingesting new routes from neighbors, running path selection, and potentially sending BGP advertisements in turn to their neighbors. Our simulator has been validated for correctness via the comparison of simulated routing tables after BGP convergence to those found in a collection of Quagga [25] routers arranged in identical topologies.

The topology used in the simulation is from Caida's inferred AS relationships dataset taken from February of 2015 [2]. This topology contains 49,755 unique autonomous systems. IP block to AS mappings are built using observations of routing tables taken from RouteViews [32] collected at the same time as the AS relationship data. All Routers were configured to obey common routing best practices, including valley free routing policies [15, 18], honoring of hybrid AS relationships and inferred partial transit agreements, and aliasing of sibling ASes. Some measurement studies, most recently Anwar *et al.* [13] have pointed out that routes observed in the wild occasionally do not always conform to valley free routing policies. In their measurements, they found

**1** Our simulator code is currently publicly available, but the link is withheld due to blind submission. If accepted a direct link to the simulator source code will be provided.

that when sibling relationships between ASes, hybrid AS relationships, and partial transit relationships [20] are considered a majority of routes, 64.7%, can be correctly predicted using no valley policies.

**Traffic Model.** In order to measure economic losses, we simulate traffic flow through the Internet using the pre and post RAD attack paths. To achieve this, we needed an Internet scale model of where traffic originates from, where its destination is, and how much of it there is. We based our model on existing work, specifically that of Gill *et al.* [19], supported by the measurements of Labovitz *et al.* [27], the World Bank [11], PeeringDB [9], and Sandvine [5]. This model takes into account both the random host to host traffic distributed across Internet hosts and traffic which is concentrated at extremely large content providers such as Netflix and Google, so called "super ASes."

Simulated traffic units fall into one of two different buckets. Traffic in the first bucket is marked as "host to host" traffic, meaning that it neither originates from, nor is addressed to, large scale content providers. The second bucket of traffic represents traffic that originates from large scale content providers and CDNs. Measurement studies have shown that the overwhelming bulk of this traffic flows from the CDNs to end hosts. In order to build a model of relative traffic ratios between these two types of we consulted the Sandvine Global Internet Phenomena Report [5], which provides a region by region breakdown of an average user's Internet traffic. The percentage of traffic from large scale content providers ranged from 30.0% percent for users in the African region, up to 67.5% percent for users in the North American region. The breakdown of the most popular sources of content by region were provided by Sandvine as well. These sources generally included Youtube/Google, Netflix and Facebook. Traffic not attributed to a specific destination was spread evenly between several other large CDN providers noted by previous measurement studies: Microsoft Azure, Akamai, Limelight, and Baidu. In order to model which ASes have CDN nodes locally, we referenced Netflix's public documentation for how it selects ASes to host content providing servers [7].

In order to establish the relative values of traffic leaving and entering ASes three data sets were combined. Sandvine provides the amount of bandwidth consumption from an "average" user in various regions. This information was combine with the World Bank's estimation of the number of Internet users in each country to get relative inbound and outbound bandwidth on a per nation state basis. In order to assign that bandwidth to ASes, we first assigned each AS to the nation state it primarily resides in. We then consulted PeeringDB, which is a system that allows ASes to advertise their willingness to peer with other ASes. ASes which elect to participate in PeeringDB have the ability to optionally disclose the average amount of inbound and outbound bandwidth from their AS that peers should expect. Of the roughly 49,000 ASes which exist in our topology just over 6,000 report bandwidth estimates. In order to establish relative bandwidth values between all ASes a Random Subspace classifier was trained based on AS features including AS degree, size of customer cone, primary country of operation and size of IP space advertised. The resulting classifier had a correlation coefficient of 0.71. Nation state bandwidth was divided proportionally between member ASes based on this estimate of inbound/outbound bandwidth to peers.

Our experiments focus on **billable traffic**, rather than total traffic. Billable traffic is every unit of traffic an AS either receives or sends to a customer AS. We focus on this traffic since each unit of billable traffic directly corresponds to *revenue* generated by the transit AS, a result of the transit AS being paid by the associated customer for carrying the traffic. The amount of money a customer pays to have that traffic delivered is typically established based on peak bandwidth consumption. Intuitively, peak load should be proportional to the total traffic sent in our simulation, thus total costs will be proportional to simulated billable traffic units in our experiments. The telecom industry consulting group TeleGeography reported that total annual IP transit revenue in 2014 was 4.9 billion US dollars [6], allowing us to compute a ratio between simulated traffic units and annual revenue of $1.66652 \times 10^{-20}$ USD of annual revenue per simulated traffic unit.

**Resistors.** We have selected four different resistors as examples. A brief summary of their properties can be seen in Table 1. **Brazil** is the largest in terms of member ASes, but the second smallest in terms of unique ASes it connects to. Our selection of Brazil was motivated by recent fallout over the NSA QUANTUM program. It is not far fetched, given current proposals [1], that Brazil would elect to act as a resistor with relation to surveillance related TMBs. **China** has a highly developed Internet infrastructure and has demonstrated a willingness to take radical steps to attack censorship circumvention systems in the past. **Germany** has expressed responses similar to Brazil's when it comes to US government surveillance. However, Germany enjoys an order of magnitude more unique points of connectivity with the rest of the world, and 24 times the number of unique providers compared to Brazil. **Iran** represents the least well connected, and consequently weakest, of our resistors.

| Resistor | Member ASes | Adjacent ASes | Provider ASes |
|---|---|---|---|
| Brazil | 3140 | 162 | 57 |
| China | 255 | 292 | 59 |
| Germany | 1792 | 1372 | 142 |
| Iran | 399 | 21 | 14 |

**Table 1.** A comparison of relevant statistics for example resistors featured in simulations.

**Deployment Selection.** In this work we focus on two types of deployments. First, we consider a **targeted deployment**, where the placement of TMBs is done with the goal of maximizing the amount of traffic from the resistor which travels across tainted paths. Censorship circumvention systems are examples of TMBs that might be deployed in such a manner. For example, one could envision a theoretical deployment of decoy routers targeting a repressive nation state such as China. Targeted surveillance is another form of TMB that might utilize this deployment strategy. In order to select which ASes to deploy to, our originator first infers what BGP paths the resistor is utilizing and weights those paths based on an estimation (or measurement) of the amount of traffic utilizing that path. Potential deployers are then scored based on the weights of the routes they appear on. The candidate AS with the largest score is then selected as a deployer, all routes which are tainted by that deployer are removed from the set of resistor routes, and the other candidate deployers are re-scored against the resistor routes which remain clean. This process is repeated until the desired number of deployers has been selected. It is worth noting that in both this deployment and the following deployment we did not limit our originator to selecting only ASes that reside in one nation, instead we allowed the originator to deploy to any AS outside of the resistor. For all of our resistors, a deployment of 10 TMBs resulted in more than 60% of the traffic leaving the resistor initially utilizing a tainted path. When deployments were larger than 40 TMBs all resistors saw 80% of their traffic traveling through at least one TMB. Figure 14, found in the Appendix, shows the percentage of traffic leaving our resistors that utilizes a tainted path initially when targeted deployments are constructed in this manner for each resistor.

We also consider a **global deployment** where the originator has the goal of maximizing the amount of traffic passing over at least one TMB regardless of the traffic's source or destination. This type of deployment would be utilized by mass surveillance programs. Selection of deployers for the global deployment occurs in the same manner as it does in the case of a targeted deployment with one change. Rather than only considering paths utilized by the resistor, all paths from all ASes are considered. This means that, unlike the targeted deployment where there is a different deployment for each resistor, there is only a single unique global deployment. This global deployment resulted in slightly less traffic from our resistors crossing at least one TMB, but still resulted in 80% of resistor traffic utilizing tainted paths when at least 40 TMBs are deployed. The percentage of traffic using tainted paths initially against our global deployment is shown in Figure 15, found in the Appendix.

# 4 Reducing Resistor Costs

It is natural to ask if there is some way to reduce the costs placed on resistors while not crippling their ability to avoid tainted paths. In this section, we examine how a resistor can launch less costly RAD attacks by altering where it considers the existence of TMBs in the BGP route selection process. We will evaluate the strength of these attacks by examining how much traffic they can move away from deployers compared to the original RAD attack.

## 4.1 Levels of Resistor Strength

The BGP route selection process allows for multiple locations where a resistor can insert logic that takes into account TMBs. In the original RAD attack TMBs were taken into account before local preference, essentially adding a rule of "Prefer clean path" to the decision process. Additionally, a rule was added to the advertisement decision process forcing an AS to advertise *all* best paths to an adjacent AS if that AS was also a resistor. Prior work [23] pointed out that resistors will suffer increased costs as a result of launching the original RAD attack. First, the change to route advertisement policy would force some non-transit ASes to act as transit providers, carrying traffic for other ASes. Second, by considering path cleanness before local preference, ASes might experience increased transit costs, as some destinations reachable via routes learned from customers would shift to non-tainted routes learned from providers. Next, since path cleanness is also considered before AS level path length, the resistor might select some routes that have longer AS level paths, possibly resulting in lower Quality of Service. Lastly, the resistor will be forced to increase the traffic load on certain links in the topology, possibly requiring the upgrading of the physical network infrastructure.

| | Original | Local Pref | Path Length | Tiebreak |
|---|---|---|---|---|
| **Transit Conversion** | Yes | No | No | No |
| **Increased Cost** | Yes | Yes | No | No |
| **Increase Path Length** | Yes | Yes | Yes | No |
| **Increased Link Load** | Yes | Reduced | Reduced | Reduced |

**Table 2.** A comparison of the costs incurred by a resistor using various strategies. The original Routing Around Decoys (RAD) attack and the proposed resistors we evaluated are shown.

In this work we present three alternative resistor strategies which seek to reduce these resistor costs. Each of these strategies is successively less costly than the last, but also less able to find non-tainted paths. A summary of the reduced costs of each resistor strategy can be found in Table 2.

First, a **Local Preference Resistor** removes the additional advertisement logic, leaving route selection the same. By allowing ASes to use normal advertisement rules, non-transit ASes will never be forced to become transit ASes. However, the resistors lose the guarantee that if at least one resistor has a clean path all resistors will have a clean path.

A **Path Length Resistor** alters when clean paths are preferred over tainted ones. Instead of doing this before Local-Pref, we could do it after Local-Pref but before AS path length. This would mean that in addition to the cost reductions of the Local Preference Resistor, an AS would never find itself in the position of using routes from a provider in preference to routes from a customer, preventing the resistor from experiencing increased transit costs.

The **Tiebreak Resistor** takes this a step further moving the check for TMBs to the end of the route selection process, directly before tie-breaking. This would, in addition to the gains of the previous two resistor strategies, eliminate selecting longer AS level paths.

## 4.2 Evaluation of New Resistors

In order for the resistor to inflict economic harm on deployers, it must be able to successfully transition traffic from tainted paths to clean paths, a process we refer to as *deflecting traffic.* If our resistor is able to successfully deflect traffic away from deployers, the deployers will suffer economically. The originator of the TMBs will in turn have to defray the costs suffered by deployers. It should be noted that unlike Schuchard et al.'s original RAD attack, in this scenario the resistor does not need to find clean paths to *all* destinations. Instead, the resistor needs to replace a *sufficient* number of tainted paths with clean ones to create noticeable economic harm to the deployers in an effort to incentivize them to remove the TMBs. A discussion of how this new strategy works against Decoy Routers and other systems where the existence of a single tainted path is sufficient for the system to function is done in Section 6.

In this section, we cover how much tainted traffic our resistors were able to deflect away from deployers, and what the consequences were for path Quality of Service. Financial costs to the resistor will be analyzed with the costs to the originator of the TMBs in Section 5. We simulated all four of our sample resistors, utilizing each of our newly proposed RAD strategies and the original RAD strategy, against both targeted and global deployments of various sizes.

### 4.2.1 Ability to Deflect Traffic

The fraction of traffic which originally crossed deployers that was successfully deflected by each RAD attack can be seen in Figure 1 and Figure 2 for targeted and global deployments respectively. As can be quickly seen from the figures, the more costly RAD attacks are more capable of deflecting traffic away from deployers in all deployment scenarios. However, even the weakest of the new RAD attacks manages to deflect some amount of traffic in all scenarios except targeted deployment against Iran.

The more well connected resistors (China, Brazil, and Germany) are able to deflect traffic to varying degrees against both deployment models using all of the RAD strategies outlined in Section 4.1. In the case of China and Brazil the Tiebreak strategy is quite weak, approaching zero deflected traffic, while Germany still manages to deflect between 5% and 22% of tainted traffic using this no cost strategy. The Path Length strategy, which is the strongest financially free strategy for the resistors, can deflect a large amount of traffic at smaller deployment sizes, but can still deflect, depending on the resistor, between 1% to 9% of tainted traffic for deployments of 100 ASes. Resistors gain between a 100% and 200% increase in deflected traffic on average by stepping up to the Local Preference strategy in exchange for monetary costs (see Section **??**). The same level of gains are seen moving from the Local Preference strategy to the original RAD attack with the exception of Brazil, which massively benefits from using the original RAD attack. This is a result of the majority of small Brazilian ASes multi-homing themselves with a geographically diverse collection of ASes located outside Brazil.

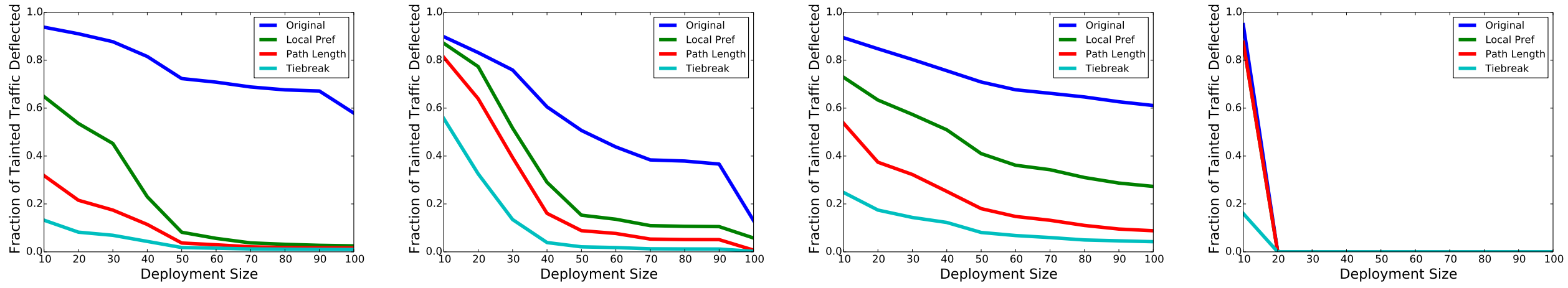

**(a)** Brazil **(b)** China **(c)** Germany **(d)** Iran

**Fig. 1.** Measured capability of our three lower cost RAD strategies, along with the original RAD attack to deflect network traffic from tainted paths, resulting from deployments targeted toward each resistor, to clean paths. Note in Figure 1d that at deployment sizes of 20 or greater Iran is completely surrounded.

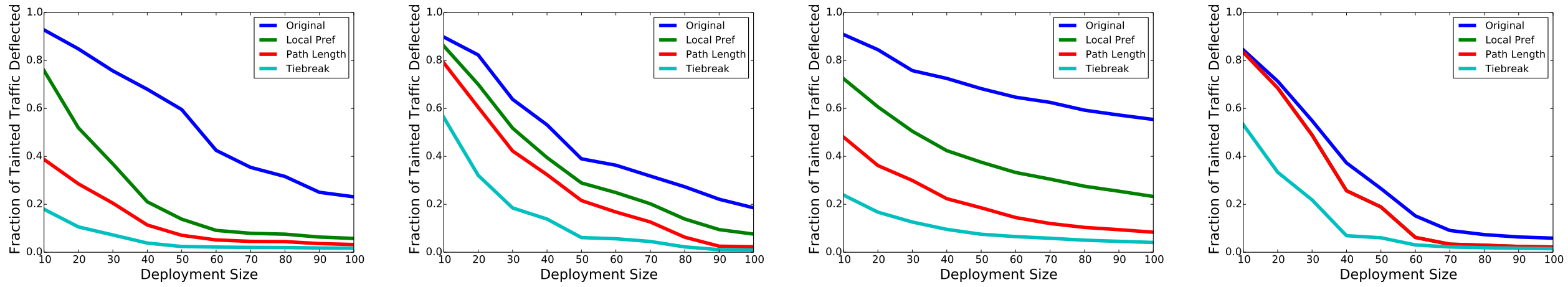

**(a)** Brazil **(b)** China **(c)** Germany **(d)** Iran

**Fig. 2.** Simulated ability for our collection or RAD strategies to deflect traffic in the face of a deployment of TMBs which aims to maximize the total traffic observed, regardless of source or destination.

Iran is unable to deflect traffic against targeted deployments of size 20 or greater since they completely surround Iran. How Iran, or any other resistor which finds itself in this scenario, is able to inflict economic harm on the deployers will be covered in Section 5.3. When considering a global deployment, even with its small pool of available alternate routes, Iran is still able to deflect traffic away from deployers. For the other resistors, the global deployment is slightly more resilient to RAD attacks compared to targeted deployments. Iran also has the only instance of two of our RAD strategies being essentially equivalent. The Local Preference and Path Length strategies return nearly identical results against a global deployment, a result of limited path choices being available to the resistor.

### 4.2.2 Quality of Service

Three out of the four tested RAD attack strategies can result in increased AS level path lengths. AS level path length, while not a direct measure of latency or path reliability, has been utilized by prior work [23] in this area as a rough measure of path quality of service. For the sake of comparison we include measurements of mean path length increase for paths which were changed during the various RAD attacks against global deployments in Figure 3. Results are similar for targeted deployments and can be seen in Figure 16 in the Appendix.

In all cases, Germany sees a mean path length increase of less than 1 AS. Brazil is the same with the exception of the original RAD attack against a targeted deployment, which has a mean increase of 1.5 ASes. China sees an increase in these same ranges for the Local Preference and Path Length strategies, but sees a mean increase of roughly 2 hops for the original RAD attack when deployment sizes are large. Iran sees no path length increases for targeted deployments as the resistor does not change paths when no clean paths can be found. Confirming that which was evident by definition, the Tiebreak strategy results in no path length changes in all cases.

More accurate measurements of path quality are difficult to achieve. We attempted to reproduce the latency experiments of Houmansadr et al.from [23], but were unable to produce estimates for the latency of even 0.1% of the changed routes using the method and data set described in [23]. Accurately estimating the latency of unused BGP routes seems to be an interesting question that is beyond the scope of this work.

Increased traffic load across a subset of links in the topology is an unavoidable consequence of altering the route decision making process. However, the relative magnitude of those increases can be mitigated by using lower cost strategies. Figure 4 shows the median relative link load increase, and Figure 5 shows the 90th percentile of relative link load increase as a fraction of normal load, only links experiencing a load increase were considered. Note that, unlike other plots in this section, these graphs are clustered via resistor strategy rather

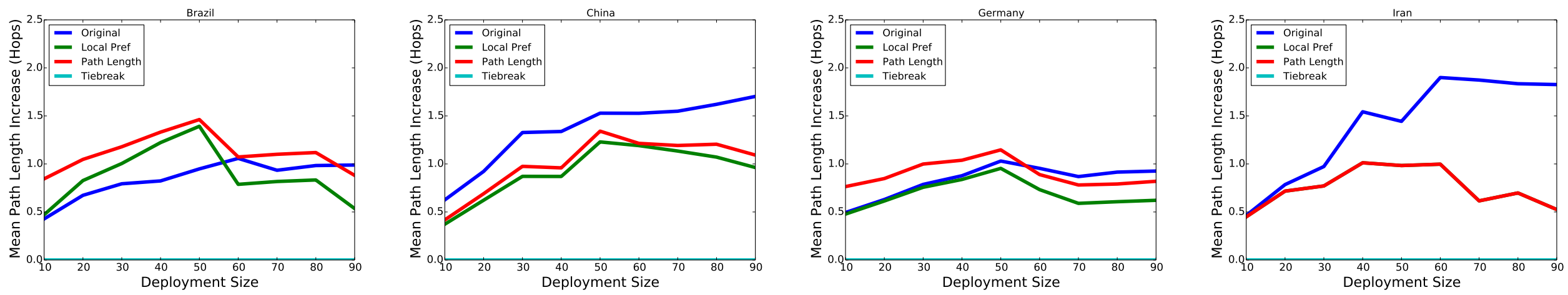

**(a)** Brazil **(b)** China **(c)** Germany **(d)** Iran

**Fig. 3.** Mean AS level path length increase see attempting to route around a global deployment.

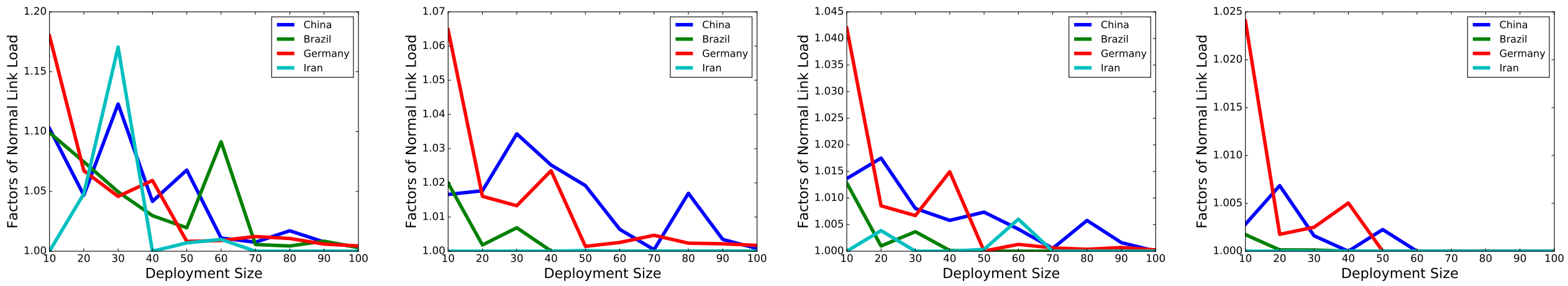

**(a)** Original **(b)** Local Pref **(c)** Path Length **(d)** Tiebreak

**Fig. 4.** Median increase in link load when considering links which must carry additional traffic as a result of resistor reaction to a global deployment. With the exception of the original RAD attack, no strategy results in more than a single digit percentage increase.

than resistor. The original RAD attack in all cases results in the largest link load increases of any strategy, a natural result of utilizing customer links for transit purposes. The other three strategies result in, with one exception, a median relative link increase that never exceeds a 5% increase in link load, meaning that the majority of links are likely not impacted from a performance perspective. The 90th percentile shows that, for smaller deployments, some links experience a doubling or tripling of traffic load. Recent measurements [17] have shown that connection points between ASes are generally less than 50% utilized. It is likely in some extreme scenarios exist where additional infrastructure would need to be built, but those are the overwhelming minority of cases. The actual monetary cost to the resistor to build this infrastructure is dependent on a large number of factors and outside the scope of this paper. One phenomena that might mitigate these financial costs is the existence of large amounts of fiber connections which are either under utilized or not currently utilized at all, so called Dark Fiber [4], which could be tapped to provide additional capacity. Additionally, many of the links which see large increases exist inside of Internet Exchange Points, or IXPs, where adding capacity, from a technical perspective, is very straightforward.

# 5 Deployer Costs

In order for TMBs to be useful, deployers must be transit providers, a type of AS that generates revenue by carrying traffic to and from their customer ASes. While transit revenue is obviously not the only way ASes generate revenue in this work we focus exclusively on transit revenue. As covered in Section 3, the amount of revenue a transit AS generates from a customer is proportional to the volume of traffic handled for that customer. Since the RAD attack directs the resistor's data away from paths that cross deployers, by extension the attack also directs the revenue that is generated by carrying this data away from the deployers.

This observation opens up a new strategy for the resistor. While the original RAD attack model focused on directly undermining the availability of TMBs, our new attack instead focuses on inflicting economic damage on deployers with the goal of incentivizing them to remove the TMBs. In this attack model our resistor is, in the short term, willing to allow traffic to cross TMBs. The resistor instead focuses on the long term goal of inflicting economic losses so severe on deployers that the originator lacks the resources to defray these losses, and the only rational decision deployers can make is to remove the TMBs. The rest of this section will measure how large the economic harm is from the traffic deflection results presented in Section 4, expand upon those attacks by taking steps to influence traffic going *to* the resistor, and examine the added incentives to defect from the ranks of deployers.

## 5.1 Direct Costs of Deployment

Using our simulator, we examine the RAD attacks of Section 4 in terms of deployer revenue losses. To quan-

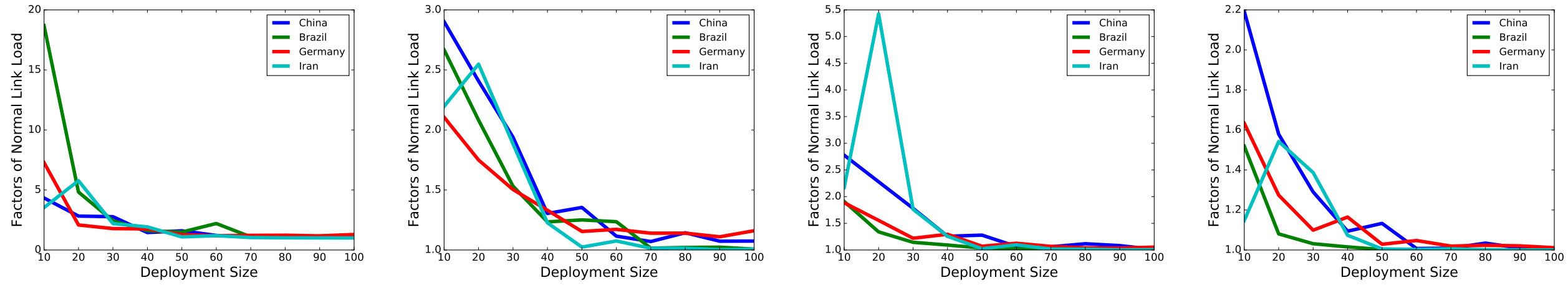

**(a)** Original **(b)** Local Pref **(c)** Path Length **(d)** Tiebreak

**Fig. 5.** The 90th percentile of link increases for our resistors employing various strategies. These increases result from reaction to a global deployment.

tify these losses, we compute the amount of revenue made by deployers before and after resistors attempt to route around them. In keeping with the concept that the resistor will not disconnect itself from any destination, during all of our simulations if a resistor can not find a clean path utilizing its chosen RAD strategy, a tainted path is utilized instead. The originator will need to reimburse the deployers for revenue lost as a result of resistor actions, and the sum of these reimbursements represents the originator's annual *cost of deployment*.

Figure 6 shows the originator's cost of deployment for targeted deployments against our resistors, and Figure 7 shows the cost of global deployments. We see that the *annual* economic damage routing capable adversaries can inflict, even using low cost RAD strategies, ranges from hundreds of thousands of dollars to tens of millions of dollars. To put those numbers in perspective, consider the United States Broadcasting Board of Governor's Internet Anti-Censorship division, the branch of the US government with the expressed task of acting as the originator for systems like decoy routing. The *total* budget in 2015 for this organization was 15 million dollars [29]. **Thus, depending on resistor strategy, the annual cost of decoy routing deployment would equal or exceed its total operating budget.**

As would be expected from the results of Section 4.2.1, More aggressive RAD strategies result in higher costs of deployment. For example, when looking at a targeted deployer being resisted by China (Figure 6b), switching from a Path Length strategy to a Local Preference strategy results in an average of 38% increase in the cost of deployment and shifting from Local Preference to the original RAD strategy gives an average improvement of 81%. Other trends from traffic deflection graphs presented in Section 4 apply to these graphs as well. For example, Brazil sees a massive increase in capabilities when switching to the original RAD attack.

The raw volume of traffic a resistor generates, in addition to what fraction of that can be deflected, now plays a role in how well our resistors can influence the cost of deployment. For example, looking back to Figure 1, Germany can deflect a higher percentage of its tainted traffic away from deployers. However, when we compare the resulting costs of deployment for Brazil, Figure 6a, to Germany, Figure 6c, we see that for many sizes of deployment, the cost of deploying against Brazil is actually larger than the cost of deploying against Germany. While Germany can find more clean paths, Brazil's directly controlled volume of traffic is larger than Germany's, meaning that the clean paths it finds can be more economically damaging to deployers. Additionally, Brazil's customers tend to have fewer choices compared to ASes that are customers of Germany, meaning that while Germany has a large customer cone, the smaller customer cone of Brazil is more likely to utilize Brazil's post-RAD attack routes.

An interesting phenomena to note is that after a certain size, larger deployments are actually *less* expensive than smaller deployments. This is a direct result of resistors being less able to find clean paths. The extreme case of this is the targeted deployment against Iran (Figure 6d) where the RAD related costs fall to zero. However, the willingness to preferentially route using clean paths results in a Prisoner's dilemma like game, creating strong incentives for deployers to remove TMBs. This game will be introduced in Section 5.3.

Two of the RAD attack strategies, the original RAD attack and the Local Preference, inflict economic damages on the resistor as well as the deployers. Both of the these strategies can force resistor ASes to utilize provider routes over customer routes. In addition, the original RAD attack can force customer ASes to provide transit for their providers. We can measure these costs as well utilizing our simulator. Figure 8 shows these resistor costs for both strategies against targeted and global deployments. We can see from the Figures that the original RAD attack is strictly more costly than a Local Preference strategy.

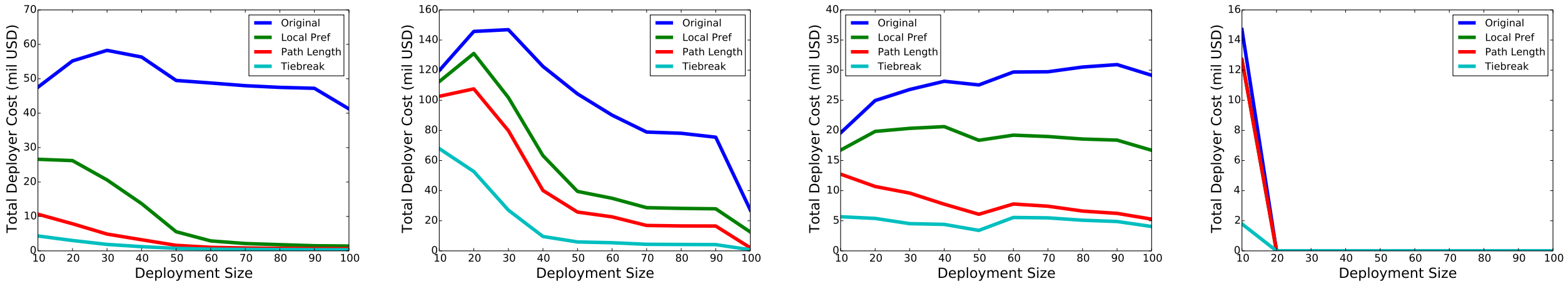

**(a)** Brazil **(b)** China **(c)** Germany **(d)** Iran

**Fig. 6.** The cost of deployments targeted against specific resistors as a function of deployment size. This is an annual cost that would have to be paid by the originator to defray lost revenue experienced by deployers.

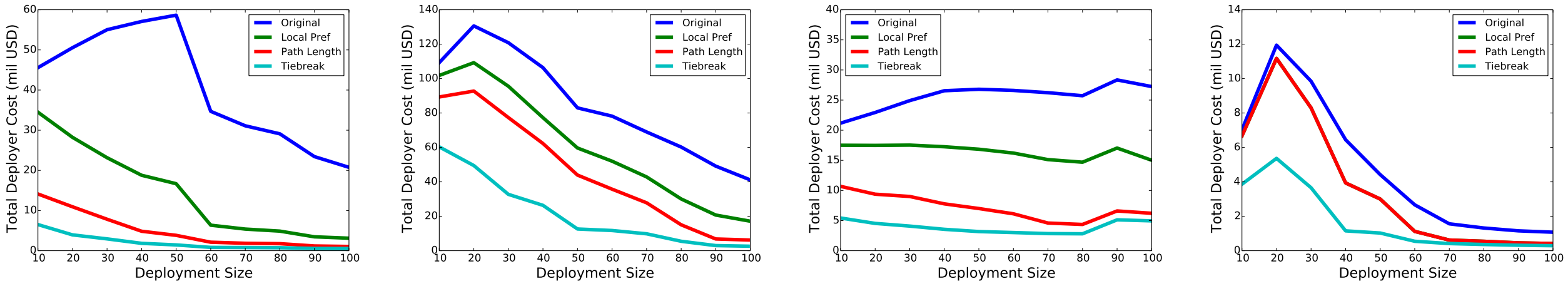

**(a)** Brazil **(b)** China **(c)** Germany **(d)** Iran

**Fig. 7.** The annual cost of a deployment maximizing global coverage of traffic when reacted to by each of the resistors.

## 5.2 Impacting Incoming Traffic

The costs of deployment presented in Figure 6 and Figure 7 were the result of resistors controlling traffic *leaving* their ASes, but what about the traffic addressed *to* their ASes? While BGP places the final decision about path selection into the hands of the routers forwarding traffic, there are techniques that give the destination AS a measure of control over what routes are used to reach their IP blocks. BGP hole punching is one such traffic engineering technique. BGP allows ASes to advertise sub-blocks of an already advertised block of IP addresses, treating these sub-blocks as a different destination. We call these routes to sub-blocks of existing destinations *hole punched routes*. When a router attempts forward a packet, it will use the path for the most specific (i.e. smallest) IP block that contains the destination IP address. In this way operators can specify different path properties for a sub-block of IP addresses through the use of hole punched routes.

Resistors can adapt this traffic engineering technique to cause other ASes to avoid sending traffic bound for resistors through tainted ASes when able. The basic idea is for resistors to generate and propagate a set of hole punched routes which are guaranteed to not cross any deployers. When an AS outside of a resistor forwards a packet to a resistor, if the AS has one of these hole punched routes, it will always use it, causing traffic that might have once traveled over a tainted path to always use a clean path.

To do this, resistor ASes partition each block of IP addresses into two sub-blocks. The resistor then advertises three paths: the original path and a path for each of the two sub blocks. The original path will spread through the Internet normally and will be used in all cases when ASes do not have a hole punched path. This ensures that the resistor faces *no connectivity loss*. The hole punched paths differ from the original path in one key way: when the resistor advertises the path after his ASN, he adds all known deployer ASes to the AS path, then adds his ASN again. The resistor can do this since no AS path integrity protection is provided in BGP. The hole punched paths will propagate through the network, but they will be rejected by deployer AS routers as a result of BGP loop detection. This causes the hole punched paths to grow *around* deployer ASes. The fact that the deployer ASes appear in the path is irrelevant for packet forwarding, since they appear after the destination AS. This technique ensures that if an AS has at least one clean path to a resistor, it will always have a clean hole punched route. We call this technique **Fraudulent Route Reverse Poisoning**, or FRRP.

We re-ran our experiments with the addition of FRRP for three deployment strategies: the original RAD attack, Local Preference, and Path Length. The Tiebreak strategy was not used in combination with FRRP since any resistor using the Tiebreak strategy has a goal of ensuring no increase in AS level path lengths. As shown in Figure 19 and Figure 20, found in the Appendix, FRRP can result in a small increase in the AS level path length for routes traveling into the resistor.

Figure 9 and Figure 10 show the costs of deployment for targeted and global deployments respectively

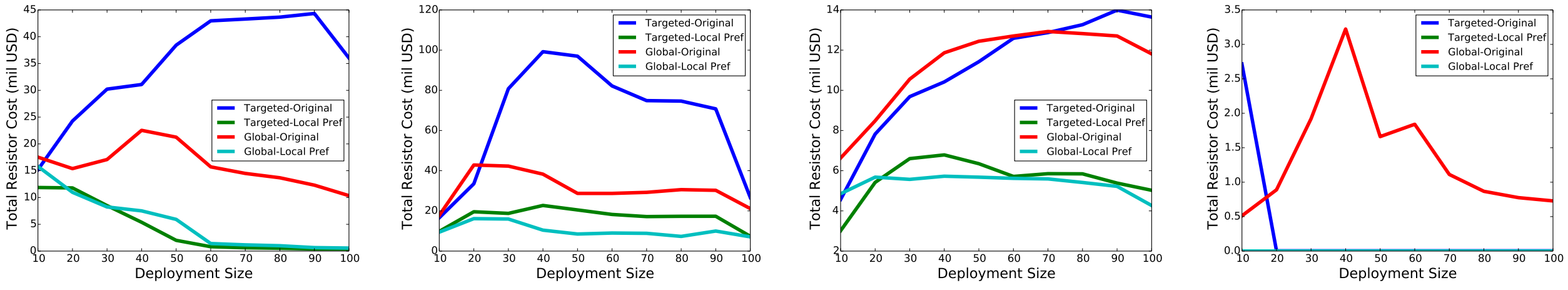

**(a)** Brazil **(b)** China **(c)** Germany **(d)** Iran

**Fig. 8.** The annual costs experienced by resistors when using the original RAD attack and the Local Preference strategy. These costs arise from having to reimburse resistor ASes for transiting traffic for their providers and added transit costs that come from migrating from customer learned paths to provider learned paths.

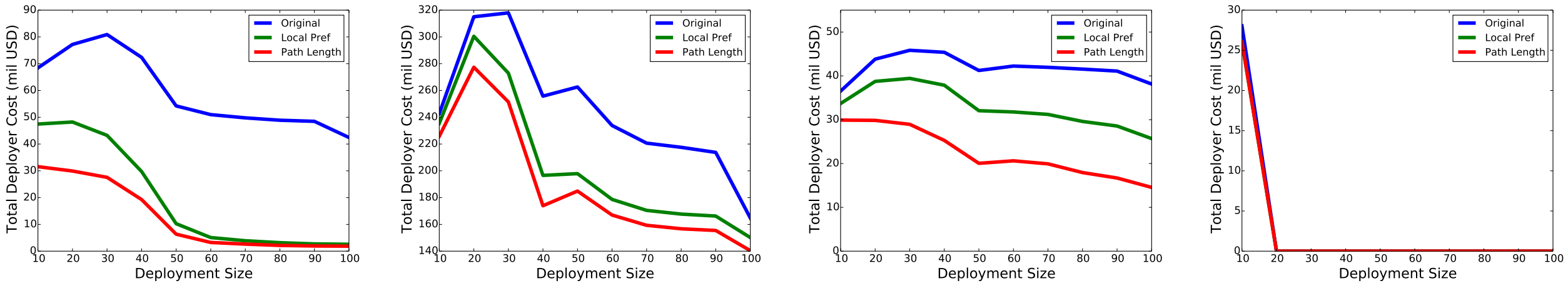

**(a)** Brazil **(b)** China **(c)** Germany **(d)** Iran

**Fig. 9.** The annual cost of targeted deployments against our resistors when both a RAD strategy and FRRP is used. Tiebreak strategy is not shown since FRRP poisoning slightly increases path length, and so is incompatible with a Tiebreak strategy.

where FRRP is employed in addition to a RAD strategy. **FRRP increases the cost of deployment by more than 40 percent in all scenarios.** In many cases the improvements are much larger, for example China can roughly double the economic losses inflicted on deployers by utilizing FRRP.

FRRP is made possible by the ability of the resistor to falsify the existence of deployers in the AS level path. This falsification can be done because of the lack of AS path integrity protection provided by BGP. Even the Resource Public Key Infrastructure (RPKI) [24], a partially deployed system attempting to provide integrity for route *origination*, provides no protection for the integrity of the path itself since the last ASN, the originator, is the correct ASN. BGPSec [14] would provide AS level path integrity, preventing the FRRP routes from being considered valid. In this case, resistors could not rely on loop detection at the deployers to guarantee that a hole punched path does not include any deployers, defeating the scheme. While BGPSec is not currently deployed, and thus not an immediate issue for a resistor, we describe a weaker version of reverse poisoning, *Selective Advertisement Reverse Poisoning*, or SelARP, which does not require falsifying path information, and would consequently still function if BGPSec were to be deployed at some future date.

SelARP takes advantage of how BGP paths are propagated through the topology. In BGP, routers will decide *if* they should advertise a route to a neighbor based on "Valley Free Routing" principles [18], meaning that advertising a path to a single neighbor does not guarantee that the route will be propagated to every AS in a topology. Deployers can therefore, by carefully selecting which neighbors they advertise the hole punched routes to, avoid allowing hole punched routes to propagate to deployers. Paradoxically, in some instances advertising a hole punched path to a neighbor which would result in a deployer later in the topology seeing the route is the optimal decision for the resistor, if prior to encountering the deployer the hole punched path manages to steal away traffic from a *different* deployer. Because the hole punched paths are not advertised to all neighbors, the guarantee that we had previously of an AS which has at least one clean path to a resistor always having at least one clean hole punched route no longer applies.

The problem of selecting which subset of neighbors to advertise the hole punched path to such that it maximizes the amount of return traffic avoiding deployers appears to reduce to the set covering problem, which is NP-hard. A greedy estimator can be built to provide a lower performance bound. This greedy estimator considers every neighbor that does not have a hole punched route advertised to it, tentatively advertises the hole punched path to it, and measures the number of ASes (weighted by IP addresses) that have a clean hole punched route after the path is propagated by BGP. If at least one neighbor increases the number of ASes with a clean hole-punched path, the neighbor which results in the most such ASes is then added to the set of neighbors who receive the hole punched path. The topology is

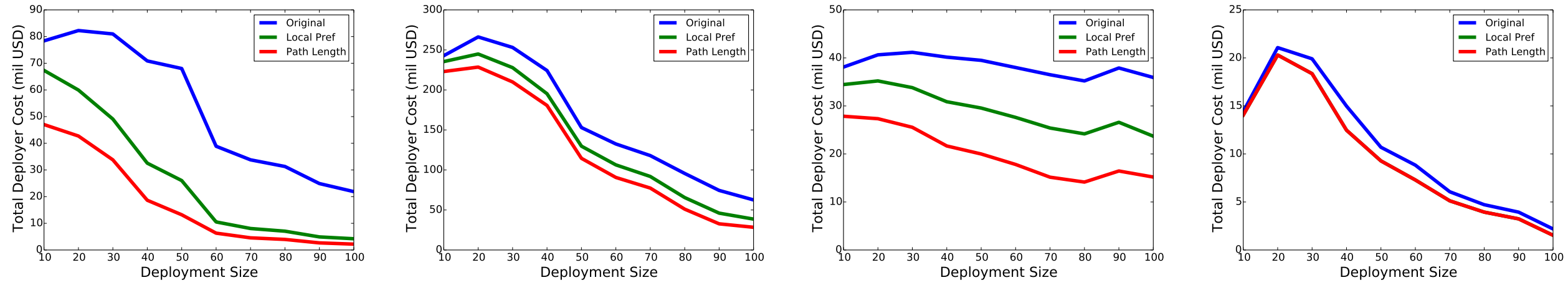

**(a)** Brazil **(b)** China **(c)** Germany **(d)** Iran

**Fig. 10.** The annual cost of a global deployment when resisted by both a RAD attack and FRRP.

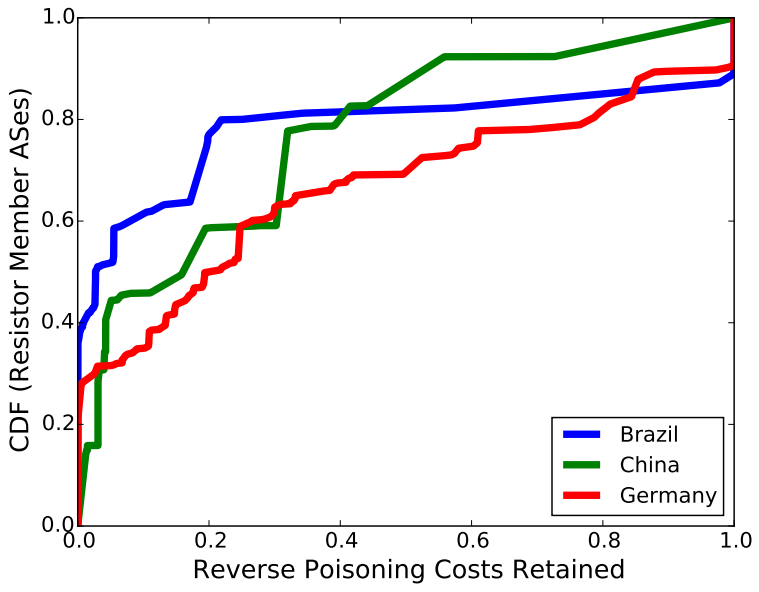

**Fig. 11.** CDF of the fraction of reverse poisoning costs that are retained when SelARP is used. This test is done against targeted deployments of 30 ASes.

then permanently updated with the hole punched route advertised to this neighbor. The process is repeated until no more such neighbors can be found. This algorithm is given in detail in the appendix.

We evaluated the ability of SelARP to influence traffic in the face of a targeted deployment. **While SelARP is weaker, it still has the ability to cause traffic which would previously traverse a deployer to instead travel a clean path.** The CDF of the fraction of incoming traffic, previously deflected via FRRP, which each resistor AS manages to still deflect away from deployers by SelARP is shown fin Figure 11 for our sample resistors. As can be seen, for the three tested resistors, the majority of member ASes are still able to deflect some amount of incoming traffic using SelARP. For Brazil and Germany nearly 20% of ASes have identical reverse poisoning capabilities when using SelARP as they do while utilizing FRRP. More refined methods for selecting which neighbors to advertise hole punched routes to, along with additional techniques such as making some hole punched routes appear less attractive by artificially inflating path length might improve these results and is left for future work.

## 5.3 Deployer Defection

If we revisit our traffic deflection graphs, Figure 1 and Figure 2, we see that a large fraction of traffic has no clean path after resistor reaction. An interesting scenario occurs when there are multiple deployers who present the resistor with a path to one of these destinations. One of those ASes will have the best (i.e. utilized) path, however resistors prefer clean paths over tainted ones. If one of the deployers who is providing a sub-optimal path to the destination removed its TMBs, an action referred to as **defection**, its path would now be the preferred path by virtue of being the only clean one. By doing this, a defecting deployer steals billable traffic away from other deployers. This means there are actually not one, but two costs experienced by deployers when they agree to host TMBs. First, there is the previously discussed loss in billable traffic that results from paths they once provided migrating to clean ASes. Second, there is an opportunity cost associated with the billable traffic they could gain by defecting and replacing other deployers as a best path provider, what we term *defection traffic.*

**When deployers force resistors to defray these opportunity costs, the cost of deployment increases by more than an order of magnitude.** Figure 12 shows the measured cost of targeted deployment when both the direct cost and the defection opportunity costs are considered. The original RAD attack, Local Preference, and Path Length strategies, all augmented with FRRP, are considered, along with the Tiebreak strategy using no FRRP. These four combinations represent a range of resistor costs ranging from the willingness to invest several million dollars to compensate member ASes down to no added cost.

The original RAD attack and Local Preference strategy (both augmented with FRRP) represent the strongest resistors. These resistors are exceptionally powerful, inflicting losses in the hundreds of millions of dollars, in the case of Brazil approaching 1 billion dollars, and in the case of China exceeding 2 billion dollars of annual cost to the originator. This amount of economic harm is large, even when considering powerful originators such as spy agencies. For example, **China would be able to require the NSA to pay out approximately** 20% **-** 30% **of its total budget [33] to**

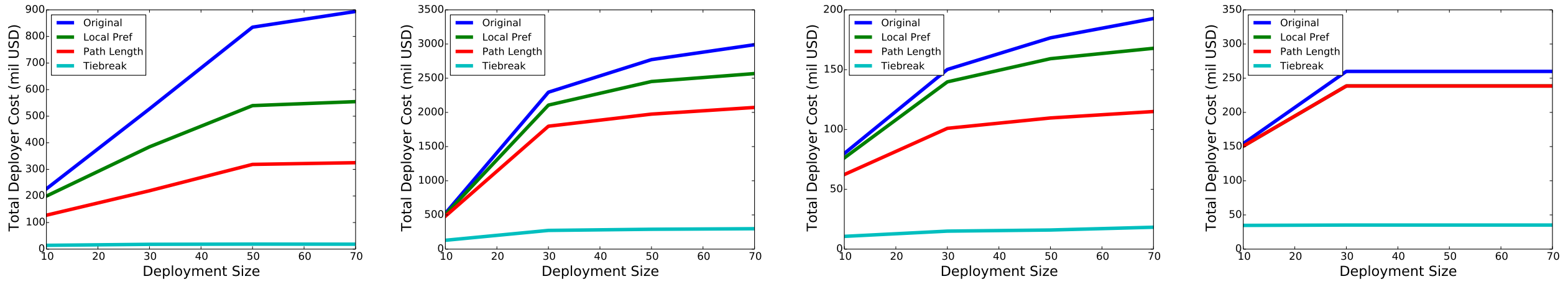

**(a)** Brazil **(b)** China **(c)** Germany **(d)** Iran

**Fig. 12.** The cost of targeted deployments, including defection costs, when a RAD attack and FRRP is utilized. Note that unlike the prior costs, when defection is considered, Iran is a viable resistor.

**cover the cost of a large deployment.** The financially free strategies can also inflict substantial losses. For all resistors, the Path Length with FRRP strategy yields costs of deployment at or exceeding roughly 100 million dollars annually. The free attack of simply Tiebreaking based on path cleanness imposes annual costs of deployment ranging from 15 million dollars for Germany to 290 million dollars for China, in all cases equaling or exceeding the total budget of the Broadcasting Board of Governor's Internet Anti-Censorship division and other free speech advocacy organizations.

Two key differences should be noted when comparing the costs of deployment including defection with those ignoring defection. First, larger deployments are no longer less costly compared to smaller deployments. This is a result of the incentive for deployers to defect *increasing* as more and more resistor traffic needs a clean path. The second phenomenon, which is closely linked to the first, is the emergence of Iran, shown in Figure 12d, as a strong resistor. As covered previously, the targeted deployment rapidly surrounds Iran with deployers, which means that all RAD attacks fail. However it also means that nearly *all* of Iran's traffic is up for grabs for any adjacent AS that elects to defect. This demonstrates the difficulty faced by the originator strategy of "blockading" a target. While establishing the blockade may be simple, maintaining it is costly.

## 6 Discussion

**Alternative Deployment Strategies.** Instead of a nation state acting as a resistor, it could alternatively act as a deployer. For example a government could have all ASes inside of a country deploy TMBs. This would of course trivially prevent clean paths from being found to any destination inside the country. On the surface this appears to be an effective deployment strategy, as resistors can not find alternative paths to any destinations inside the nation. However, the interconnected nature of the Internet means that quite often a large fraction of traffic flowing through a transit AS is bound for a nation *other* than the one the transit AS resides in. While it is true that all of the traffic destined for delivery inside the deployer nation state can not be diverted away from deployers without connectivity loss, traffic passing through the deployer state can. Therefore deployer nations risk the loss of this traffic due to resistor action. Figure 13 shows a CDF of the fraction of transit traffic flowing through transit ASes in our simulator that is bound for a destination in a different nation than the one the transit AS resides in. As can been seen, for approximately 80% of transit ASes more than a third of all transit traffic is bound for a destination abroad. Fully two fifths of transit ASes see half or more of their traffic heading to foreign destinations.

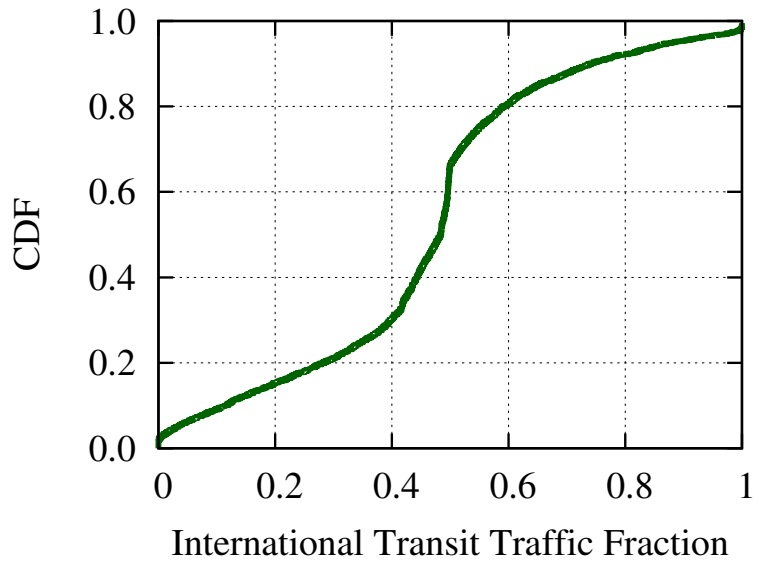

**Fig. 13.** A CDF of the fraction of transit traffic bound for a destination that is international relative to the transit AS.

**When Originators are nation states.** In some of the motivating examples we highlighted TMBs that conduct surveillance for a nation state. This scenario presents multiple additional challenges to our resistors. To begin with, the budgets of spy agencies, as covered in Section 5, are far larger than that of NGOs, on the order of billions of dollars. However, some resistors, for example China, can generate deployment costs on the order of several billion dollars annually against global deployments. It should also be noted that if multiple resistors launch economically focused RAD attacks the losses inflicted on deployers only increase. Given that one of our motivating examples is a set of alleged TMBs deployed by the NSA, it is interesting to consider a deployment scenario similar to the one just discussed, where the United States is ringed with TMBs. In this scenario, rather than focus on a single nation state resisting we consider a coalition of the top 5% of ASes, ranked by degree of connectivity, acting as a resistor. TMBs are deployed to all ASes which primarily operate in the United States that have a link to an AS which resides in another nation. The deployment costs, both direct and those including defection costs are shown in Table 3. The first thing to note is that even though no clean paths exist into the United States, there are still direct losses on the order of half a billion dollars annually as a result of some international transit traffic no longer traversing through the United States. In addition, we see that defection pressure increases deployment costs by nearly an order of magnitude for all strategies except Tiebreak.

| | Direct Costs | Defection Costs |
|---|---|---|
| **Legacy** | **529.7 mil USD** | **5017 mil USD** |
| **Local Pref** | **496.5 mil USD** | **4150 mil USD** |
| **Path Len** | **422.7 mil USD** | **3036 mil USD** |
| **Tiebreak** | **174.2 mil USD** | **871 mil USD** |

**Table 3.** Direct and opportunity costs for ASes belonging to the United States when a "ring" deployment around the US is resisted by a coalition of 5% of all ASes.

When the originator is a nation state other complications for our resistor's attack model exist. For example there is the question of whether or not deployers are allowed to act as rational economic entities. If the deployers are legally compelled to host TMBs without compensation then the originator would not bear any of the cost of our resistor's actions. However, it is important to note that our resistor's actions are still functional, as they provide a strong economic incentive, which did not exist prior to the resistor actions, for the deployers to push back against the forced deployment of TMBs to their network, which is the resistor's goal in the first place. An additional complication might be the surreptitious deployment of TMBs, in such a case a "deployer" may be unaware of the location or even presence of TMBs on its network, making defection improbable – although with resistor detection and action, high enough opportunity cost may motivate the deployer AS to discover these facts.

**Finding deployers of TMBs.** We note that for several examples of TMBs, it is essentially trivial to identify deploying ASes. Schuchard *et al.* showed how, in the original RAD attack, deployers can be identified by using the decoy routing protocol itself: the efficacy of the protocol depends on users being able to signal a covert destination to the TMB. Thus, a given AS path can be tested using this signal; if the covert destination is retrieved, then the path includes a deployer. Repeating this process with each AS path available to the resistor allows the resistor to identify the deployers using an intersection attack. Given a known keyword, similar processes have been used to identify ASes that manipulate traffic by keyword filtering [41], ad injection [30], and DNS injection [12]. Work has even recently been presented which has had some amount of success detecting the QUANTUM INSERT infrastructure [21].

More subtle traffic manipulation such as traffic shaping and protocol differentiation might require probabilistic inferences by the resistor [16, 37], leading to the possibility of classification errors. In this case, we note that since the new attacks we discuss have little cost to the resistor, *false positives have limited consequence.* Let $C$ be the cost our attack incurs on a deployer, $\alpha$ be the true positive rate, and $\psi$ be the false positive rate. Then a TMB deployer will have expected cost $\alpha C$ under our attack, and a non-deployer will have expected cost $\psi C$. Thus as long as $\alpha > \psi$ the attacks will disincentivize deployment; however, compared to manipulations that can be detected perfectly, the damages will be reduced by a factor of $(\alpha - \psi)$. Fully passive surveillance systems present the most challenging scenario for our resistor, however non-technological detection, for example through disclosures similar to those of Edward Snowden could be sufficient for resistors to launch a RAD attack against all member ASes of a given nation state to place pressure on the host government.

**TMBs Which Require Total Disconnection** Readers familiar with Decoy Routers will have observed that they function so long as *any* tainted path exists from a given resistor. In this case our economic resistor will return to the policy of refusing to send any traffic to destinations which lack a clean path, however the economic resistor's position is strengthened via economic pressure on deployers. By increasing the cost on the originator, whole more than likely has a fixed budget to spend on TMB deployment, the resistor can reduce the size of deployment the originator can field. This in turn means that there will be fewer end destinations the resistor will need to cut itself off from when preventing any traffic from utilizing tainted paths.

**Modeling limitations.** We note that there are some limitations to our cost model, which may tend to overstate the opportunity costs imposed by defection. Additionally, variations in pricing and contractual obligations to other organizations could also complicate the decision to defect. Without more detailed information about the pricing policies and exact traffic volumes of individual ISPs, it is difficult to measure the impact of a resistor's actions with perfect accuracy. The models put forward in this paper exist to illustrate the concept of an economically based attack against availability rather predict exact losses to the last dollar.

The other limitation of our cost model is the fact that only a single round of deployer incentives and defection is examined. Our defection costs present a maximum opportunity cost, as it assumes in each case that only one AS elects to defect. If two ASes simultaneously elected to defect it is possible that the amount of additional transit traffic they attract is lower than either expects; a result of the two ASes splitting the gains. However the opposite direction is also true, two or more deployers might form a *defection coalition*, creating clean paths which previously had multiple deployers. In this way the single defector scenario is not necessarily a worst-case scenario. An interesting question is what are originator costs when multiple rounds of deployment and defection are conducted? Additionally, preferential routing by our resistor could incentivize other ASes to build links to resistor ASes in an effort to capture traffic from the resistor which currently has no clean path. We feel that examining these scenarios is an exciting direction for future work.

A general concern is the accuracy of our simulated paths. Because of the nature of resistor behavior, specifically some of the models violating common operator practices for route selection and export specification, we can not take an approach similar to RAPTOR [36] where we only use routes observed in the wild. Even so, we should generally find a majority of the routes our simulator selects appearing in the wild at some point in time. This in fact is true, we find that the majority of our simulated paths are seen in the wild, specifically our simulator aligns with Anwar *et al.*'s results that approximately two thirds of routes should be correctly predicted. A full, table of the fraction of pre and post RAD attack paths for each of our resistor that we can observe in use is found in the Appendix. Another complication that can result in errors in simulated paths is the existence of peering links between ASes which are difficult to detect from the outside. In order to test that our model is insensitive to the existence of such links we tested our attacks on topologies with additional peering links added to the topology. We found that in the majority of cases the addition of more peering links made our routing capable adversary stronger. This is shown in Figure 17 and Figure 18 in the Appendix. We elected to use only the observable AS topology since it resulted in the weakest adversarial model.

**Detection of reverse poisoning.** We note that as described, the reverse poisoning attack can be detected by an AS that monitors all incoming route advertisements. In principle, it might be possible to develop a routing policy that detects this and counters the poisoning by re-advertising the aggregate route for each sub-block. This policy would carry substantial risks of introducing routing cycles and other instabilities in the BGP infrastructure, and thus does not seem to be carried out in practice. Moreover, non-lying reverse poisoning would be far harder to detect. Additionally, ASes detecting non-lying reverse poisoning could only react to it by falsely claiming they too have the hole punched routes, something they would be unable to accomplish in the presence of BGPSec.

# 7 Conclusion

In this work we demonstrated how a collection of like minded ASes can fight back against devices that observe and manipulate their traffic from privileged positions in the Internet. First, we showed that the original RAD attack can be adjusted to reduce costs on resistors. Through simulation we show that these adjusted RAD attacks are still capable of deflecting non-negligible traffic away from deployers in two likely to occur deployment models. Second, we demonstrated that a RAD attack inflicts economic damages on deployers, which would inhibit a resistor's ability to find willing ASes to deploy TMBs. Our experiments demonstrate that by directing profitable traffic away from deployers, resistors can financially incentivize ASes to remove TMBs from their network. By making known the fact that clean paths are preferentially selected, resistors can create a Prisoner's dilemma-like game which causes economic pressure to be exerted, even if the resistor can not currently find clean paths to destinations.

The economic attack we developed creates many interesting questions for future work. First, what happens when a more long term game is played between resistors and the originator? Our examination of the pressure to defect looks at only a single round game. What changes when the game is run over multiple rounds? What does the clean path fraction look like when ASes make decisions to deploy or defect over time? What are the actual economic costs when multiple deployers defect simultaneously?

There is also a large amount of work that can be done examining deployment strategies. Are there deployment strategies that work more efficiently when economic concerns are considered? Additionally, we can consider how to build resistor coalition. Certain collections of ASes might be found which function together as a powerful economic adversary by virtue of their combined connectivity and customer cone.

## Acknowledgements

This work was supported by NSF grants 1314637 and 1223421. The first author was supported by a University of Minnesota Doctoral Dissertation Fellowship.

## References

[1] Brazil, europe plan undersea cable to skirt u.s. spying. http://www.reuters.com/article/2014/02/24/us-eu-brazil-idUSBREA1N0PL20140224. Accessed: 26 February 2014.

[2] CAIDA AS relationship dataset. http://www.caida.org/data/active/as-relationships/index.xml.

[3] Cisco visual networking index: Forecast and methodology, 2012 to 2017. http://www.cisco.com/c/en/us/solutions/collateral/service-provider/ip-ngn-ip-next-generation-network/white_paper_c11-481360.html. Accessed: 15 May 2014.

[4] Dark fiber community. http://www.darkfibercommunity.com/.

[5] Global internet phenomena report. https://www.sandvine.com/trends/global-internet-phenomena/.

[6] Ip transit forecase. https://www.telegeography.com/research-services/ip-transit-forecast-service/index.html. Accessed: 26 August 2015.

[7] Netflix open connect program. https://openconnect.netflix.com/en/.

[8] The NSA uses powerful toolbox in effort to spy on global networks. http://www.spiegel.de/international/world/the-nsa-uses-powerful-toolbox-in-effort-to-spy-on-global-networks-a-940969.html. Accessed: 15 May 2014.

[9] Peeringdb. https://www.peeringdb.com/.

[10] Ripe ris raw data. https://www.ripe.net/analyse/internet-measurements/routing-information-service-ris/ris-raw-data.

[11] World bank global indicators. http://data.worldbank.org/indicator/.

[12] Anonymous. The collateral damage of internet censorship by dns injection. *SIGCOMM Comput. Commun. Rev.*, 42(3):21–27, June 2012.

[13] R. Anwar, H. Niaz, D. Choffnes, I. Cunha, P. Gill, and E. Katz-Bassett. Investigating interdomain routing policies in the wild. In *Proceedings of the 2015 ACM Conference on Internet Measurement Conference*, pages 71–77. ACM, 2015.

[14] S. Bellovin, R. Bush, and D. Ward. Security Requirements for BGP Path Validation. RFC 7353 (Informational), Aug. 2014.

[15] M. Caesar and J. Rexford. Bgp routing policies in isp networks. *Network, IEEE*, 19(6):5–11, nov.–dec. 2005.

[16] M. Dischinger, M. Marcon, S. Guha, P. K. Gummadi, R. Mahajan, and S. Saroiu. Glasnost: Enabling end users to detect traffic differentiation. In *NSDI*, pages 405–418, 2010.

[17] N. Feamster. An unprecedented look into utilization at internet interconnection points. https://freedom-to-tinker.com/blog/feamster/the-interconnection-measurement-project-revealing-utilization-at-internet-interconnection-points/. Accessed: 11 March 2016.

[18] L. Gao and J. Rexford. Stable internet routing without global coordination. *IEEE/ACM Transactions on Networking (TON)*, 9(6):681–692, 2001.

[19] P. Gill, M. Schapira, and S. Goldberg. Let the market drive deployment: A strategy for transitioning to bgp security. In *ACM SIGCOMM Computer Communication Review*, volume 41, pages 14–25. ACM, 2011.

[20] V. Giotsas, M. Luckie, B. Huffaker, et al. Inferring complex as relationships. In *Proceedings of the 2014 Conference on Internet Measurement Conference*, pages 23–30. ACM, 2014.

[21] L. Haagsma. Deep dive into quantum insert. http://blog.fox-it.com/2015/04/20/deep-dive-into-quantum-insert/. Accessed: 15 May 2015.

[22] A. Houmansadr, G. T. Nguyen, M. Caesar, and N. Borisov. Cirripede: circumvention infrastructure using router redirection with plausible deniability. In *Proceedings of the 18th ACM Conference on Computer and Communications Security (CCS)*, 2011.

[23] A. Houmansadr, E. L. Wong, and V. Shmatikov. No direction home: The true cost of routing around decoys. In *Proceedings of the 2014 Network and Distributed System Security (NDSS) Symposium*, 2014.

[24] G. Huston and G. Michaelson. Validation of Route Origination Using the Resource Certificate Public Key Infrastructure (PKI) and Route Origin Authorizations (ROAs). RFC 6483 (Informational), Feb. 2012.

[25] K. Ishiguro et al. Quagga Routing Suite. http://quagga.net/.

[26] J. Karlin, D. Ellard, A. W. Jackson, C. E. Jones, G. Lauer, D. P. Mankins, and W. T. Strayer. Decoy routing: Toward unblockable internet communication. In *Proceedings of the USENIX Workshop on Free and Open Communications on the Internet (FOCI)*, 2011.

[27] C. Labovitz, S. Iekel-Johnson, D. McPherson, J. Oberheide, and F. Jahanian. Internet inter-domain traffic. *ACM SIGCOMM Computer Communication Review*, 41(4):75–86, 2011.

[28] J. Menn. Secret contract tied nsa and security industry pioneer. http://www.reuters.com/article/2013/12/20/us-usa-security-rsa-idUSBRE9BJ1C220131220. Accessed: 16 February 2015.

[29] B. B. of Governors. Broadcasting board of governors: Internet anti-censorship fact sheet. http://www.bbg.gov/wp-content/media/2015/04/Anti-Censorship-Fact-Sheet-10182015.pdf. Accessed: 28 Aug 2015.

[30] C. Reis, S. D. Gribble, T. Kohno, and N. C. Weaver. Detecting in-flight page changes with web tripwires. *NSDI*, 8:31–44, 2008.

[31] Y. Rekhter, T. Li, and S. Hares. A Border Gateway Protocol 4 (BGP-4). RFC 4271 (Draft Standard), Jan. 2006. Updated by RFCs 6286, 6608, 6793.

[32] RouteViews. RouteViews Dataset. http://www.routeviews.org/.

[33] J. Sahadi. What the nsa costs taxpayers. http://money.cnn.com/2013/06/07/news/economy/nsa-surveillance-cost/. Accessed: 28 Aug 2015.

[34] B. Schneier. Attacking tor: how the NSA targets users' online anonymity. http://www.theguardian.com/world/2013/oct/04/tor-attacks-nsa-users-online-anonymity. Accessed: 15 May 2014.

[35] M. Schuchard, J. Geddes, C. Thompson, and N. Hopper. Routing around decoys. In *Proceedings of the 2012 ACM Conference on Computer and Communications Security*, CCS '12, pages 85–96, New York, NY, USA, 2012. ACM.

[36] Y. Sun, A. Edmundson, L. Vanbever, O. Li, J. Rexford, M. Chiang, and P. Mittal. Raptor: routing attacks on privacy in tor. In *24th USENIX Security Symposium (USENIX Security 15)*, pages 271–286, 2015.

[37] M. B. Tariq, M. Motiwala, N. Feamster, and M. Ammar. Detecting network neutrality violations with causal inference. In *Proceedings of the 5th International Conference on Emerging Networking Experiments and Technologies*, CoNEXT '09, pages 289–300, New York, NY, USA, 2009. ACM.

[38] N. Weaver. A close look at the NSA's most powerful internet attack tool. http://www.wired.com/2014/03/quantum/. Accessed: 15 May 2014.

[39] E. Wustrow, S. Wolchok, I. Goldberg, and J. A. Halderman. Telex: anticensorship in the network infrastructure. In *Proceedings of the 20th USENIX Security Symposium*, 2011.

[40] E. Wyatt. F.c.c., in a shift, backs fast lanes for web traffic. http://www.nytimes.com/2014/04/24/technology/fcc-new-net-neutrality-rules.html?smid=pl-share&_r=0. Accessed: 15 May 2014.

[41] X. Xu, Z. M. Mao, and J. A. Halderman. Internet censorship in china: Where does the filtering occur? In *Passive and Active Measurement*, pages 133–142. Springer, 2011.

# A Appendix

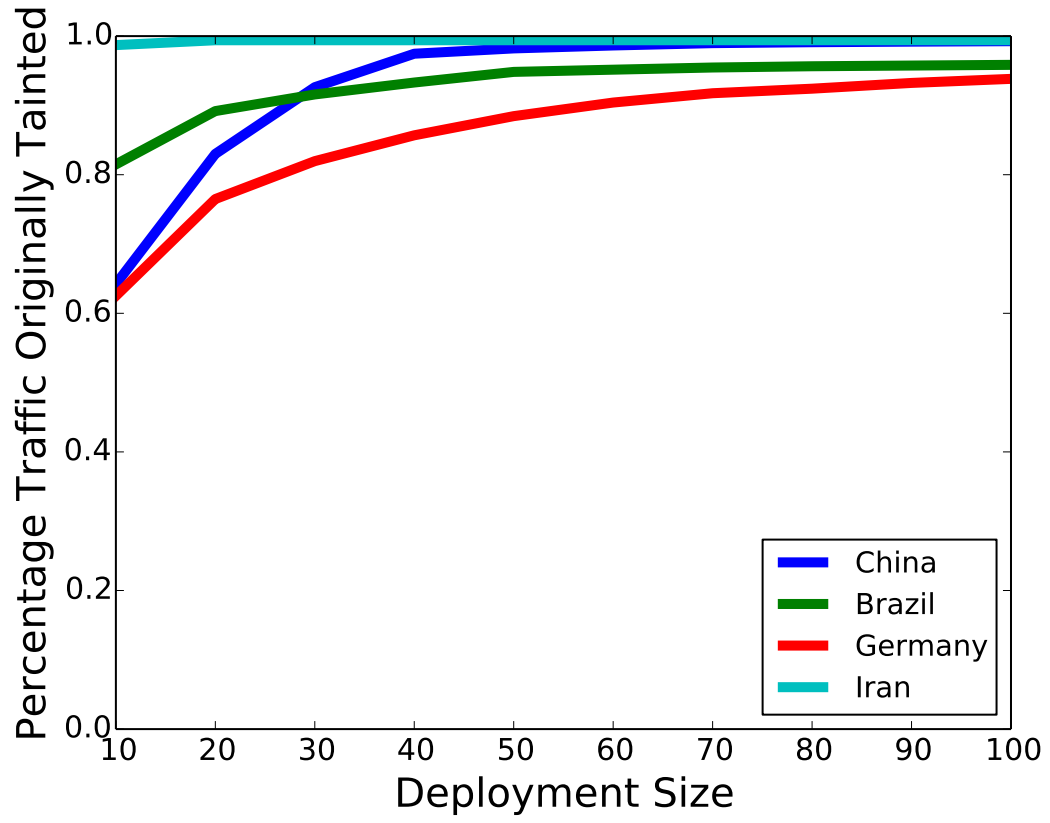

**Fig. 14.** The fraction of traffic originating inside our resistors which crosses at least one TMB placed in a targeted deployment.

**Data**: Set of neighbors
**Result**: Set of neighbors to advertise hole punch
*score* ← 0;
*AdvertiseSet* ← {};
*Unadvertised* ← *Neighbors*;
**repeat**
  *roundGain* ← 0;
  *roundChoice* ← ∅;
  **forall the** *candidates* ∈ *Unadvertised* **do**
    advertise hole punched route to *candidate*;
    *tempScore* ← current deployer losses;
    **if** *tempScore* - *score* > *roundGain* **then**
      *roundGain* ← *tempScore* − *score*;
      *roundChoice* ← *candidate*;
    **end**
    withdraw hole punched route to *candidate*;
  **end**
  advertise hole punched route to *roundChoice*;
  *score* ← *score* + *roundGain*;
  *AdvertiseSet* ← *AdvertiseSet* ∪ *roundChoice*;
  *Unadvertised* ← *Unadvertised* \ *roundChoice*;
**until** *roundGain* = 0;

**Algorithm 1:** Greedy algorithm for selecting which neighbors to advertise the hole punched route to. The algorithm does a test advertisement to each neighbor that has not been selected, and selects the neighbor which results in the highest deployer losses. Each time a neighbor is selected the hole punched route is permanently advertised to them.

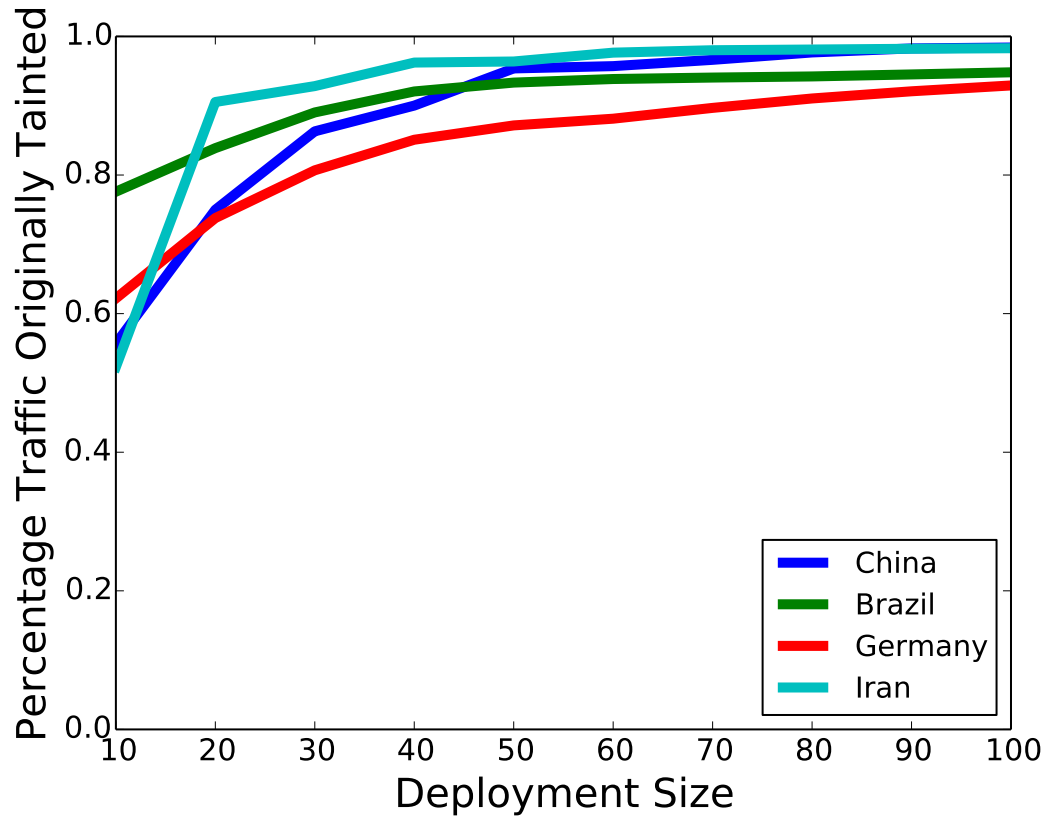

**Fig. 15.** The fraction of traffic resistor originated traffic which crosses at least one TMB deployed as part of a global deployment.

| | Initial Observed | Resistor Used Observed |
|---|---|---|
| **Brazil** | 63.2% | 46.6% |
| **China** | 62.2% | 42.2% |
| **Germany** | 56.4% | 49.1% |
| **Iran** | 45.3% | 13.4% |

**Table 4.** One way we can measure the accuracy of our topology is to compare the routes generated by our software routers to actual observed routes collected a route reflectors around the Internet, similar to the approach taken by RAPTOR [36]. We parsed information from RIPE route reflectors [10] and Route-Views peering points [32] within a three month window before and after when our inferred AS relationship dataset was generated by CAIDA. The first column is the percentage of routes prior to resistor attacks which appear in the wild. The second column is the percentage of resistor reaction routes which can be observed in the wild. The second column is lower as a result of many of the resistor strategies violating no Valley Routing.

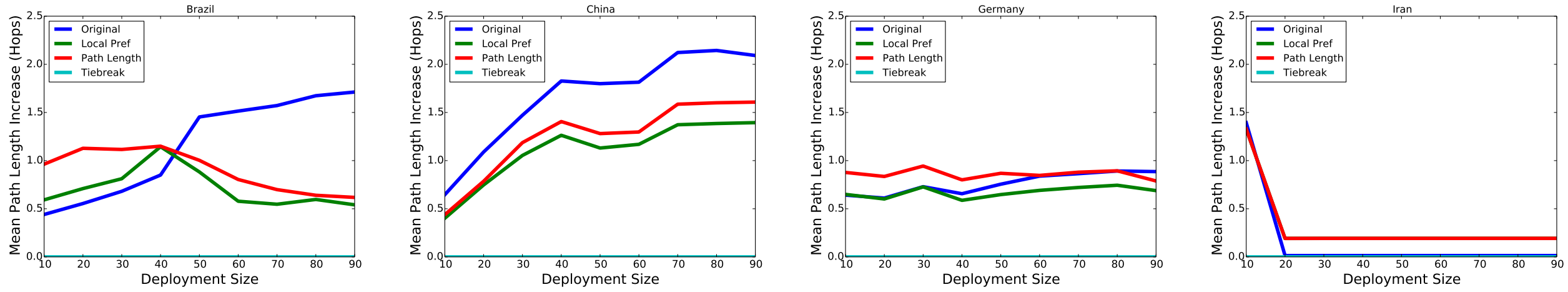

**(a)** Brazil **(b)** China **(c)** Germany **(d)** Iran

**Fig. 16.** Mean AS level path length increase for our collection of RAD strategies operating against the targeted deployment. In all cases the Tiebreak strategy sees zero path length increase.

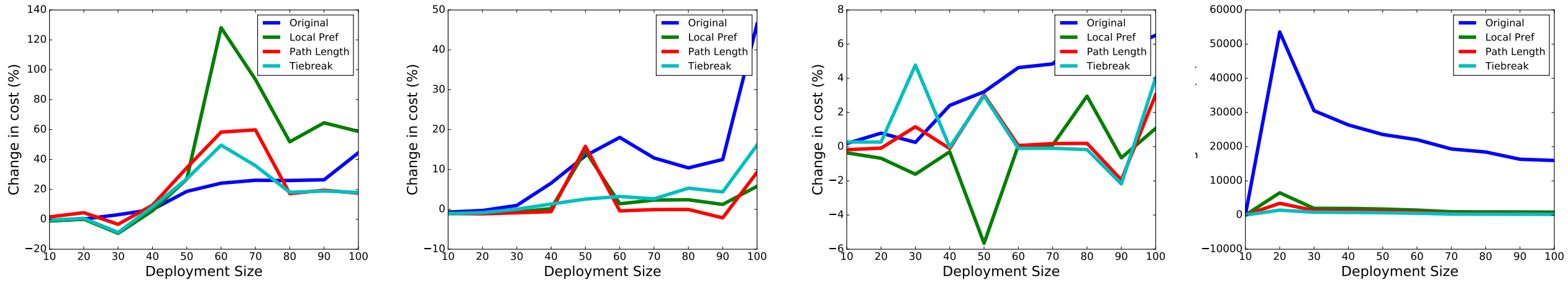

**(a)** Brazil **(b)** China **(c)** Germany **(d)** Iran

**Fig. 17.** Change in cost of deployment when our topology has an additional 10% of peering links added. Peering links are difficult to detect and may be missing from our AS topology. By adding additional peering links in nearly all cases our resistor is stronger.

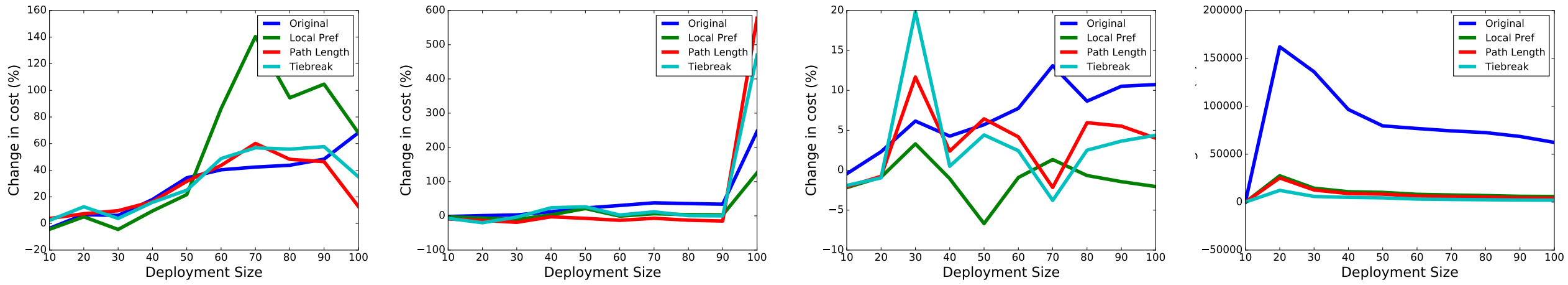

**(a)** Brazil **(b)** China **(c)** Germany **(d)** Iran

**Fig. 18.** Chane in the cost of deployment when our topology has an added 30% of peering links. Again, our resistors are stronger as a result of the added links.

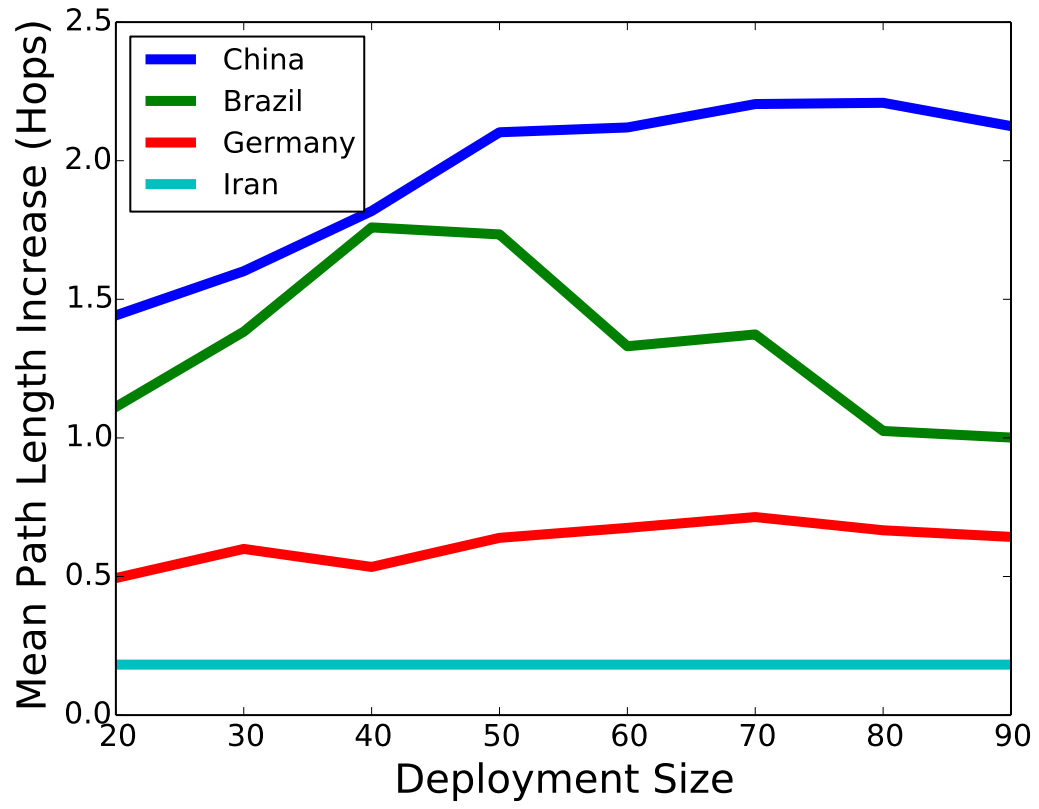

**Fig. 19.** The mean increase in path length *into* the resistors as a result of reverse poisoning against a targeted deployment.

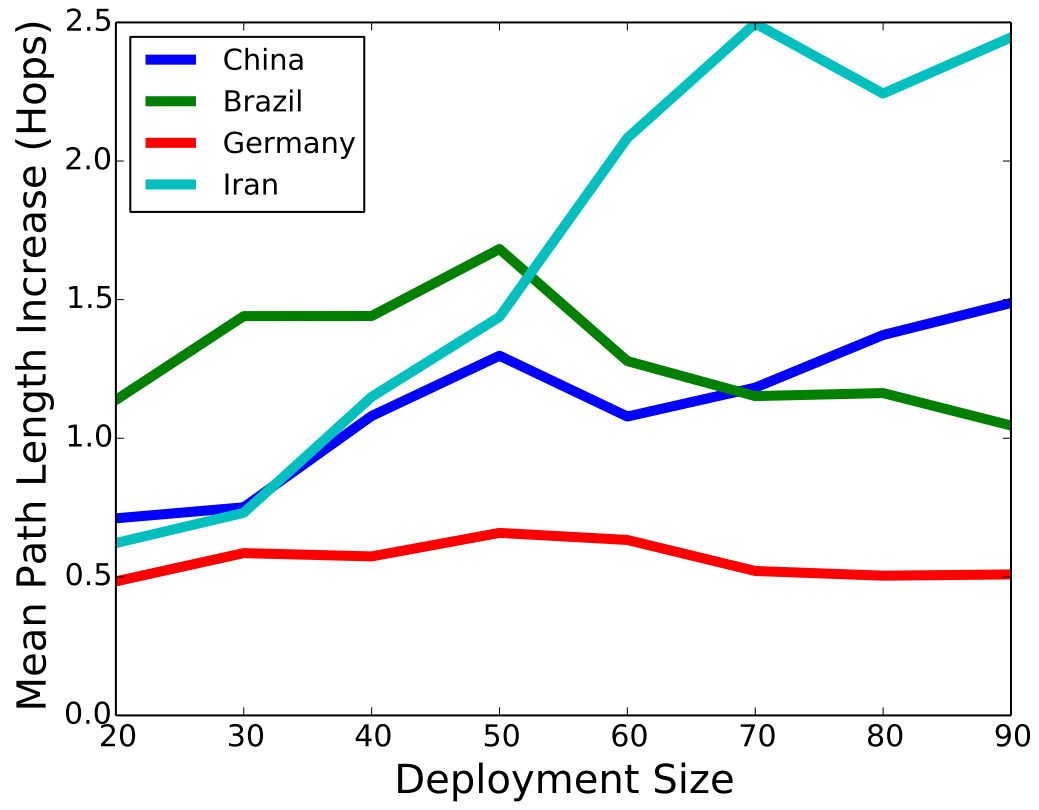

**Fig. 20.** The mean increase in path length for in-bound paths to the resistor as a result of reverse poisoning against a global deployment.